\documentclass[final,twocolumn,5p]{elsarticle}

\usepackage{amssymb}
\usepackage{siunitx}
\usepackage{booktabs}
\usepackage{IEEEtrantools}
\usepackage{amsmath}

\usepackage[figuresright]{rotating}

\DeclareMathOperator\erf{erf}

\usepackage{color}

\begin{document}
\bstctlcite{IEEEexample:BSTcontrol}
\begin{frontmatter}




\title{Energy Measurements by Means of Transition Radiation in novel Linacs}



\author[mymainaddress,mysecondaryaddress]{M. Marongiu\corref{mycorrespondingauthor}}
\cortext[mycorrespondingauthor]{Corresponding author}
\ead{marco.marongiu@uniroma.it}
\author[mythirdaddress]{M. Castellano}
\author[mythirdaddress]{E. Chiadroni}
\author[myfourthaddress]{A. Cianchi}
\author[mythirdaddress]{G. Franzini}
\author[mythirdaddress]{A. Giribono}
\author[mymainaddress,mysecondaryaddress]{A. Mostacci}
\author[mymainaddress,mysecondaryaddress]{L. Palumbo}
\author[mythirdaddress]{V. Shpakov}
\author[mythirdaddress]{A. Stella}
\author[mythirdaddress]{A. Variola}

\address[mymainaddress]{Sapienza University, Rome, Italy}
\address[mysecondaryaddress]{INFN-Roma, Rome, Italy}
\address[mythirdaddress]{LNF-INFN, Frascati, Italy}
\address[myfourthaddress]{INFN-Roma ''Tor Vergata'', Rome, Italy}
\begin{abstract}
Advanced linear accelerator design may use Optical Transition Radiation (OTR) screens to measure beam spot size; for instance, such screens are foreseen in plasma based accelerators (EuPRAXIA@SPARC\_LAB) or Compton machines (Gamma Beam Source@ELI-NP). Optical Transition Radiation angular distribution strongly depends on beam energy. Since OTR screens are typically placed in several positions along the Linac to monitor the beam envelope, one may perform a distributed energy measurement along the machine. Furthermore, a single shot energy measurement can be useful in plasma accelerators to measure shot to shot energy variations after the plasma interaction. Preliminary measurements of OTR angular distribution of about \SI{100}{MeV} electrons have been performed at the SPARC\_LAB facility. In this paper, we discuss the sensitivity of this measurement to beam divergence and others parameters, as well as the resolution required and the needed upgrades of conventional OTR diagnostics, using as an example the data collected at SPARC\_LAB.

\end{abstract}


\end{frontmatter}


\section{Introduction}
The Gamma Beam Source~\cite{Bacc2013} (GBS) machine is an advanced source of up to $\approx$\SI{20}{MeV} Gamma rays based on Compton back-scattering, i.e. collision of an intense high power laser beam and a high brightness electron beam with maximum kinetic energy of about \SI{720}{MeV}.  
The Linac will provide trains of bunches in each RF pulse, spaced by the same time interval needed to recirculate the laser pulse in a properly conceived and designed laser recirculator. Thus, the same laser pulse will collide with all the electron bunches in the RF pulse, before being dumped. The final design foresees trains of 32 electron bunches separated by \SI{16}{ns}, distributed along a \SI{0.5}{\micro\second} RF pulse, with a repetition rate of \SI{100}{Hz}. 

In a typical monitor setup, the beam is imaged via Optical Transition Radiation (OTR) or YAG screen using standard lens optics, and the recorded intensity profile is a measure of the particle beam spot. In conjunction with other accelerator components, it will also be possible to perform various measurements on the beam, namely: its energy and energy spread (with a dipole), bunch length~\cite{Fili2011} (with an RF deflector), Twiss parameters~\cite{Most2012} (by means of the quadrupole scan technique)  or in general 6D characterization on bunch phase space~\cite{Cian2015}. Such techniques are common in conventional~\cite{ferrario2013sparc_lab} and unconventional~\cite{Most2012,antici2012laser,rossi2014external} high brightness Linacs. In this paper, we refer unconventional or novel Linacs to the plasma based accelerators (both beam and laser driven) and to the GBS machine. The reason why, in our opinion, the GBS machine can be defined as a novel Linac is due to the fact that it will produce high brightness multi-bunch pulses  (bunch by bunch separation of \SI{16}{ns}) that will be accelerated by a newly designed, and not yet fully characterized, C-Band accelerating structures~\cite{alesini2017design}.  Such schemes could pose different challenges in terms of beam stability that need to be measured by the appropriate diagnostics.

Since OTR screens are typically placed in several positions along the Linac to monitor the beam envelope, one may perform a distributed energy measurement along the machine. This will be useful, for instance, during the commissioning phase of the GBS in order to verify the correct functionality of the newly designed C-Band accelerating structures~\cite{alesini2017design}, due to the fact that there are OTR screens after each accelerating module. 

Furthermore, a single shot energy measurement can be useful in plasma accelerators to measure shot to shot energy variations after the plasma interaction (i.e. EuPRAXIA@SPARC\_LAB~\cite{ferrario2017eupraxia}). In order to perform this measurement with ultra short beams (typical in plasma accelerators), one needs to take into account also the coherent OTR whose contribution is neglected in this paper. Moreover, for this type of beams, the use of dipoles to perform energy measurements could be critical, due to the high energy jitter.  

Several techniques have been proposed to measure energy of a beam with high jitter  using a spectrometer; for instance, in this study~\cite{nakamura2008broadband} the proposed configuration (with a total length of 1 meter) foreseen one dipole and 2 scintillating screens that can measure beam energy in the range from \SI{10}{MeV} to \SI{1.1}{GeV}. Other studies~\cite{sears2010high,soloviev2011two} proposed schemes with both a permanent magnet and an electromagnetic spectrometer to increase the resolution in an energy range that goes from \SI{2}{MeV} to \SI{400}{MeV}. A simpler and more compact (\SI{25}{cm}) scheme~\cite{glinec2006absolute} is based on a permanent magnet spectrometer and 1 lanex screen for low charge beam in the energy range from \SI{20}{MeV} to \SI{200}{MeV}. The technique proposed in this paper, however, cover a wide range of energies (i.e. from \SI{30}{MeV} to \SI{3}{GeV}) with a compact, cheap and already installed hardware (i.e. OTR screen, CCD sensor, lenses). Moreover, if a different range of energies or an improvement of resolution is needed, one can easily change ``in air'' optics without modifying in vacuum devices. This type of measurement meets also the requirement of having a compact Linac since it does not need any bending magnet.

This paper describes a theoretical concept of the OTR-based electron beam property measurements, followed by the experimental study using a 100-MeV class conventional accelerator (SPARC\_Lab). Conclusions and outlook are presented as well.

\section{Theory}

Optical Transition Radiation screens are widely used for beam profile measurements, as well as in ELI-GBS~\cite{marongiu2017optical,cioeta2017spot}. The radiation is emitted when a charged particle beam crosses the boundary between two media with different optical properties. For beam diagnostic purposes the visible part of the radiation is used; an observation geometry in backward direction is chosen corresponding to the reflection of virtual photons at the screen which acts as a mirror. 

The main advantages of OTR are the instantaneous emission process allowing fast single shot measurements, and the good linearity (neglecting coherent effects); indeed, the typical response time of the OTR is \SI{10}{fs}~\cite{lumpkin1998time} while for a YAG screen is \SI{70}{ns}~\cite{YAG_crytur}. The disadvantages are that the process of radiation generation is invasive, (i.e. a screen has to be inserted in the beam path and, unless a properly designed thin OTR foil is used, the beam got completely scattered when it passes through the screen), and that the radiation intensity is much lower in comparison to scintillation screens.

\begin{figure}[!htb]
   \centering
   \includegraphics*{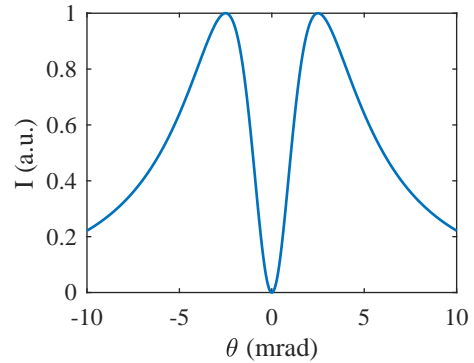}
   \caption{Horizontal profile of the OTR angular distribution of a single electron (Single Particle Function).}
   \label{fig:Ginzburg_matlab}
\end{figure}

Another advantage of the OTR is the possibility to measure the beam energy by means of observation of its angular distribution (see figure~\ref{fig:Ginzburg_matlab}); this technique has been proved feasible by many authors~\cite{ginsburg1946radiation,wartski1975thin}, also for low energy beams~\cite{castellano1995analysis}. The angular distribution can be expressed by the well known formula~\cite{ginsburg1946radiation}:
\begin{equation}
\frac{dI^2}{d\omega d\Omega}=\frac{e^2}{4\pi^3c\epsilon_0}\frac{\sin^2{\theta}}{\left(\frac{1}{\gamma^2}+\sin^2{\theta}\right)^2}R(\omega,\theta),
\label{eq:frank}
\end{equation}
where $\omega$ is the frequency, $\Omega$ is the solid angle, $I$ is the intensity of the radiation, $e$ is the electron charge, $c$ is the speed of light, $\epsilon_0$ is the vacuum permittivity and $R(\omega,\theta)$ is the reflectivity of the screen; the peak of intensity is at $\theta=1/\gamma$ with respect to the beam direction. 

Due to the beam divergence, the angular distribution of the whole beam will be different from $0$ at the center: the ratio between the minimum and the maximum intensity is related to the beam divergence. A parameter called visibility can be defined as in Eq. (\ref{eq:visibility}): in analogy with the contrast function, the measurement with the OTR angular distribution can be reliably done if the visibility parameter is greater or equal to  $0.1$~\cite{cianchi2016transverse}. 
\begin{equation}
V=\frac{I_{max}-I_{min}}{I_{max}+I_{min}}.
\label{eq:visibility}
\end{equation}
Assuming a Gaussian distribution of the divergences, the OTR angular distribution can be written as the convolution between Eq. (\ref{eq:frank}) and the Gaussian distribution as in Eq. (\ref{eq:cianchi}).
\begin{align}
 I&\propto\frac{\sqrt{\pi}\mu}{\nu}\Re{\left[\Phi(z)\left(\frac{1}{2}+\mu\nu z\right)\right]-\mu^2}, \nonumber\\
 \mu&=\frac{1}{\sqrt{2}\sigma'}, \quad \Phi(z)=\frac{1-\erf{(z)}}{\exp{[-z^2]}}, \nonumber \\  z&=\mu(\nu+i\theta), \quad \nu=\frac{1}{\gamma}, 
 \label{eq:cianchi}
 \end{align}
 where $\erf(z)$ is the complex error function and $\Re$ is the real part~\cite{cianchi2016transverse}.

As it can be seen in Eq. (\ref{eq:cianchi}), $I_{max}$ and $I_{min}$ depends on both divergence and energy of the beam. Equation~\ref{eq:visibility}, therefore, implicitly gives the range of beam energy and divergence over which this technique can be used: since for bigger energies the angular distribution narrows, the sensitivity to angular spread is higher than for low energy beams where the angular distribution is wide. For instance, for a beam energy of \SI{700}{MeV}, the divergence must be below \SI{2}{mrad}; for a beam energy of \SI{5}{GeV}, the divergence must be below \SI{0.3}{mrad}, while for a beam energy of \SI{140}{MeV}, the divergence must be below \SI{10}{mrad}.

Moreover, the beam energy has an effect on the ability of a given optic system to resolve the angular distribution, since the angular distribution narrows as the energy increases; therefore, a change of the optic system (i.e. a bigger focal length) could be necessary.

\section{Experiment}
\begin{table}[hbt]
\center{
\caption{Main beam parameters for two different working points at SPARC\_LAB. The values were measured with conventional devices and techniques (Beam current monitor for the charge, dipole for the energy and quadrupole scan for the beam divergence). The values between brackets represent the uncertainty of the measurements.}
  \label{tab:beam_spot}
\begin{tabular}{l r r}
\toprule
& \textbf{Data set 1} & \textbf{Data set 2} \\
\midrule
$E$ (\SI{}{MeV}) & 110.82 (0.07) & 123.1 (0.04) \\
$\Delta E/E$ (\SI{}{\%}) & 0.13 (0.002) & 0.06 (0.0002) \\
$Q$ (\SI{}{pC}) & 108 (3) & 120 (4) \\
$\sigma'_x$ (\SI{}{mrad}) & 0.52 (0.03) & 1.1 (0.09) \\
$\sigma'_y$ (\SI{}{mrad}) & 0.66 (0.02) & 1.04 (0.09) \\
\bottomrule
\end{tabular}
}
\end{table}
In this section we shown the application of the technique described in the previous section to the high brightness photoinjector of SPARC\_LAB; we verified the feasibility of the technique for different values of charge, energy, divergence) and with different measurement setup (i.e. single shot and time integrated measurements).

Equation~\ref{eq:cianchi} was used to retrieve the beam energy and divergence for different machine working points. The  first working point, called ``Data set 1'', was characterized by lower charge, energy and divergence with respect to the second working point, called ``Data set 2'' (see Table~\ref{tab:beam_spot}). The optic layout used to observe the OTR angular distribution was the same for the different working points and it was reported in~\cite{cianchi2016transverse}.

\begin{table}[hbt]
\center{
\caption{Beam energy and divergence measured at SPARC\_LAB for the ``Data Set 1'' working point and for 3 different configurations (Single shot, 1 second integration and 5 seconds integration). The values between brackets represent the uncertainty of the measurements.}
  \label{tab:data_set1}
\begin{tabular}{l r r r}
\toprule
\textbf{Data set 1} & $E$ (\SI{}{MeV}) & $\sigma'_x$ (\SI{}{mrad}) & $\sigma'_y$ (\SI{}{mrad}) \\
\midrule
Single Shot & 105.35 (2.04) & 0.72 (0.21) & 0.74 (0.17)\\
10 shots & 108.33 (1.53) & 0.75 (0.09) & 0.77 (0.08) \\
50 shots & 109.87 (0.55) & 0.72 (0.04) & 0.78 (0.06)\\
\bottomrule
\end{tabular}
}
\end{table}
The measurements of the first working point, in the single shot configuration, were affected by a low Signal to Noise Ratio (SNR); the  coefficient of determination of the fit  (R-square) was \SI{0.65}{} while the uncertainty was around \SI{1.9}{\%} for the energy and below \SI{30}{\%} for the divergence. 

\begin{figure}[htb]
\centering
\includegraphics*{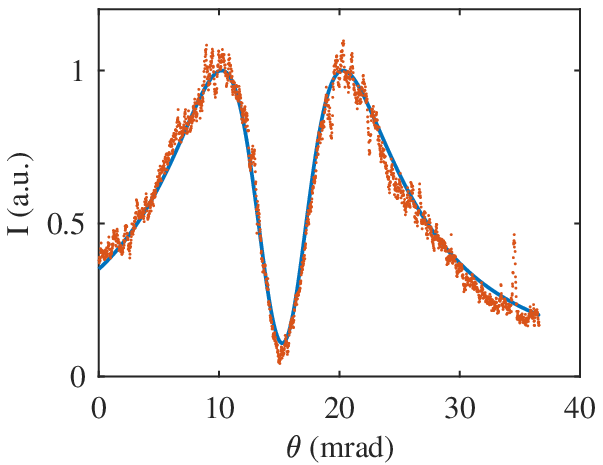}
\caption{Horizontal profile of the OTR angular distribution of a \SI{108}{pC} beam with energy of \SI{111}{MeV} and divergence of \SI{0.6}{mrad} (``Data Set 1'' in Table~\ref{tab:beam_spot}). The red dots represents the data of a \SI{5}{s} Integration measurement (the machine operates at a repetition rate of \SI{10}{Hz}), while the blue line is the fitting curve (Eq. (\ref{eq:cianchi})).}
\label{fig:5sInt}
\end{figure}

A \SI{1}{s} integration and a \SI{5}{s} integration  measurements were performed as well: the SNR  was increased,  as well as the goodness of fit. In the  \SI{1}{s} integration case, for instance, the R-square value became \SI{0.92}{} while the uncertainty became around \SI{1.4}{\%} for the energy and below \SI{12}{\%} for the divergence. The  \SI{5}{s} integration case, shown in figure~\ref{fig:5sInt}, gave an R-square value of \SI{0.97}{} while the uncertainty was around \SI{0.5}{\%} for the energy and below \SI{8}{\%} for the divergence. 

Also the accuracy of the measurement, calculated with respect to the values in table~\ref{tab:beam_spot}, increased: for the energy measurement, it went from \SI{95}{\%} of the single shot case to the \SI{99}{\%} of the \SI{5}{s} integration case (in the \SI{1}{s} integration case, the accuracy was \SI{98}{\%}). For the divergence, instead, the accuracy remained around a value of \SI{90}{\%} (see Table~\ref{tab:data_set1}).

\begin{figure}[htb]
\centering
\includegraphics*{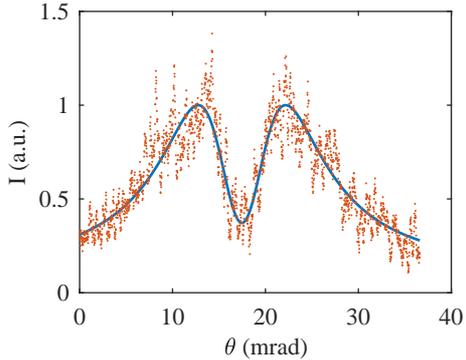}
\caption{Horizontal profile of the OTR angular distribution of a \SI{120}{pC} beam with energy of \SI{123}{MeV} and divergence of \SI{1.1}{mrad} (``Data Set 2'' in Table~\ref{tab:beam_spot}). The red dots represents the data of a single shot measurement, while the blue line is the fitting curve (Eq. (\ref{eq:cianchi})).}
\label{fig:SS}
\end{figure}

\begin{table}[hbt]
\center{
\caption{Beam energy and divergence measured at SPARC\_LAB for the ``Data Set 2'' working point and for 2 different configurations (Single shot and 1 second integration). The values between brackets represent the uncertainty of the measurements.}
  \label{tab:data_set2}
\begin{tabular}{l r r r}
\toprule
\textbf{Data set 2} & $E$ (\SI{}{MeV}) & $\sigma'_x$ (\SI{}{mrad}) & $\sigma'_y$ (\SI{}{mrad}) \\
\midrule
Single Shot & 122.13 (2.04) & 1.4 (0.1) & 1.3 (0.1)\\
10 shots & 123.66 (1.02) & 1.3 (0.05) & 1.2 (0.04) \\
\bottomrule
\end{tabular}
}
\end{table}
For the second working point, the measurements were done in the single shot configuration and with \SI{1}{s} integration; in the first case, shown in figure~\ref{fig:SS}, the  R-square value was \SI{0.82}{} while the uncertainty was around \SI{1.7}{\%} for the energy and below \SI{8}{\%} for the divergence. 

\begin{figure}[htb]
\centering
\includegraphics*{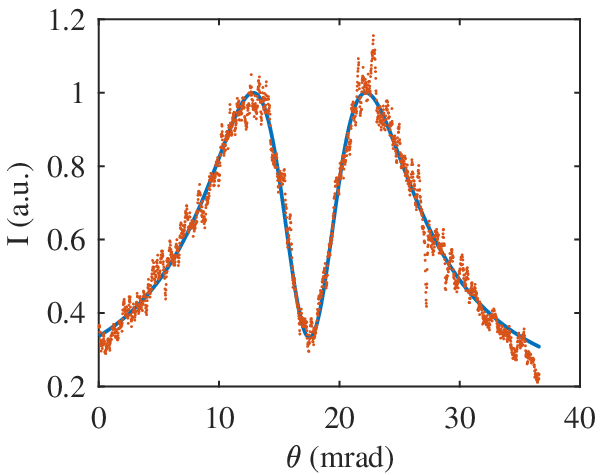}
\caption{Horizontal profile of the OTR angular distribution of a \SI{120}{pC} beam with energy of \SI{123}{MeV} and divergence of \SI{1.1}{mrad} (``Data Set 2'' in Table~\ref{tab:beam_spot}). The red dots represents the data of a \SI{1}{s} integration measurement (the machine operates at a repetition rate of \SI{10}{Hz}), while the blue line is the fitting curve (Eq. (\ref{eq:cianchi})).}
\label{fig:1sInt}
\end{figure}

In the \SI{1}{s} integration case, shown in figure~\ref{fig:1sInt}, the  R-square value was \SI{0.98}{} while the uncertainty was around \SI{0.8}{\%} for the energy and below \SI{4}{\%} for the divergence. 

The accuracy was \SI{99}{\%} for the energy and around \SI{80}{\%} for the divergence in the single shot case, and it increased to \SI{99.5}{\%} for the energy and \SI{85}{\%} for the divergence in the \SI{1}{s} integration case (see Table~\ref{tab:data_set2}).

In order to perform a distributed energy measurement along the GBS, these results were promising: since the OTR intensity is linearly dependent on the charge and, due to the fact that the GBS bunch charge is \SI{250}{pC}, this uncertainty and accuracy results are expected for a beam energy around \SI{50}{MeV}. 

Furthermore, the beam energy has an effect on the OTR intensity and on the angular spread;  the appropriate optics must be used in order to perform an accurate fit, putting enough points between the peaks and in the tails (a common rule of thumb is to acquire in the range $\theta\in [-3/\gamma:3/\gamma]$). This can be done changing the focal length (a bigger focal length implies a smaller field of view) or the sensor pixel size; in any case, the same optic system guarantees a wide range of energies (i.e. the one used in this experiment has a focal length of \SI{400}{mm} and it can measure energies between \SI{30}{MeV} and \SI{3}{GeV} but with an increased uncertainty). For lower energy (i.e. \SI{10}{MeV}), a smaller focal length must be chosen. In the latter case, the intensity is decreased and an intensifier becomes fundamental.

Finally, if a single shot measurement is needed, the uncertainty doubles with respect to the 1 second integration case both for the energy and the divergence.

\section{Conclusion and outlook}

The OTR could be a very useful diagnostic tool in order to measure the beam energy. Distributed energy measurements are foreseen especially to evaluate the performances of the accelerating structures at the ELI-GBS facility during the commissioning stage; indeed, the facility will be equipped with OTR diagnostic stations after each accelerating module. This measurement will be useful in particular for the evaluation of the newly designed C-Band structures~\cite{alesini2017design}. 

Furthermore, the energy measurement is foreseen not only for multi-bunch pulses, but also  for a single bunch of the pulse train, using  a gated camera system (i.e. Hamamatsu Orca4). In this case, the goal is to measure a single bunch within the pulse (i.e. first bunch of the first train, second bunch of the second train, etc.) and to evaluate the effects of the head bunch on the tail bunches; this could be done only in the commissioning stage, since this technique doesn't have the required resolution to measure the in-spec energy jitter shot to shot.

The energy jitter shot to shot could also be evaluated after plasma interaction if the SNR is high enough (i.e. high energy, high charge). Indeed, the data analysis shows a strong dependence of the uncertainty to the SNR; also the accuracy of the measurement is affected by the SNR. 

The main advantages of this technique are the use of diagnostics already in place (OTR screens) and its compactness since no dipole is needed.  In case of a high energy jitter, this technique does not require any tuning due to its wide range of applicability ($E_{max}/E_{min}\approx 100$). 

The experimental data shows that the uncertainty of the measurement is good enough (around 2\%) and that, in the single shot configuration, it doubles with respect to the 1 second integration case. This is useful for plasma accelerated beams (i.e. EuPRAXIA@SPARC\_LAB~\cite{ferrario2017eupraxia}).

\section*{Acknowledgments}
This work was supported by the European Union's Horizon 2020 research and innovation programme under grant agreement No. 653782.

We wish to give a special thank to the SPARC\_LAB group for their contribution with the data acquisition.
\section*{References}
\bibliographystyle{IEEEtran}
\bibliography{myPaper,OTR,Introduction,ELI,others}

\end{document}